\documentclass[a4paper,11pt]{article}
\pdfoutput=1 
\usepackage{jcappub}
\usepackage{amsmath,amssymb}
\usepackage{bm}
\usepackage{xcolor}
\usepackage{graphicx}
\usepackage{enumitem}
\usepackage{geometry}
\usepackage{xspace}
\usepackage{pdflscape}
\usepackage{placeins}
\usepackage{epstopdf}
\usepackage{comment}
\makeatletter
\gdef\@fpheader{}
\g@addto@macro\bfseries{\boldmath}
\makeatother

\newcommand{\OmegaGW}{\Omega_{\mathrm{GW}}}
\newcommand{\rhoGW}{\rho_{\mathrm{GW}}}

\let\oldsqrt\sqrt
\def\sqrt{\mathpalette\DHLhksqrt}
\def\DHLhksqrt#1#2{%
\setbox0=\hbox{$#1\oldsqrt{#2\,}$}\dimen0=\ht0
\advance\dimen0-0.2\ht0
\setbox2=\hbox{\vrule height\ht0 depth -\dimen0}%
{\box0\lower0.4pt\box2}}

\newcommand{\dd}{\mathrm{d}}

\newcommand{\sss}[1]{{\scriptscriptstyle{#1}}}
\newcommand{\boldmathsymbol}[1]{{\ensuremath{\boldsymbol{#1}}}}

\newcommand{\uPl}{\mathrm{Pl}}

\newcommand{\usssPl}{\sss{\uPl}}

\newcommand{\calH}{\mathcal{H}}

\newcommand{\Mp}{M_\usssPl}

\newcommand{\beq}{\begin{equation}}
\newcommand{\eeq}{\end{equation}}
\newcommand{\bea}{\begin{equation}\begin{aligned}}
\newcommand{\eea}{\end{aligned}\end{equation}}

\newlength{\wsingfig}
\newlength{\wdblefig}
\newlength{\wquadfig}
\newlength{\wtriplefig}
\newcommand{\Eq}[1]{Eq.~(\ref{#1})}

\newcommand{\Fig}[1]{Fig.~{\ref{#1}}}

\newcommand{\Sec}[1]{Sec.~\ref{#1}}

\newcommand{\Hc}[1]{\mathcal{H}}

\renewcommand{\Hc}{\mathcal{H}}

\def\doi{http://doi.org}

\date{today}

\title{Primordial black holes and induced gravitational waves in non-singular matter bouncing cosmology}

\author[a,b,c]{Theodoros Papanikolaou}

\author[d,]{Shreya Banerjee}
 
\author[e,f,g]{Yi-Fu Cai}

\author[a,b,h]{Salvatore Capozziello}

\author[c,f,i]{Emmanuel N. Saridakis}

\affiliation[a]{Scuola Superiore Meridionale, Largo San Marcellino 10, 80138 Napoli, Italy.}

\affiliation[b]{Istituto Nazionale di Fisica Nucleare (INFN), Sezione di Napoli, Via Cinthia 21, 80126 Napoli, Italy.}

\affiliation[c]{National Observatory of Athens, Lofos Nymfon, 11852 Athens, Greece.}

\affiliation[d]{Department of Physics,
Indian Institute of Technology (Indian School of Mines),
Dhanbad, Jharkhand-826004, India.}

\affiliation[e]{Department of Astronomy, School of Physical Sciences, University of Science and Technology of China, Hefei 230026, China.}

\affiliation[f]{CAS Key Laboratory for Researches in Galaxies and Cosmology, School of Astronomy and Space Science, University of Science and Technology of China, Hefei, Anhui 230026, China.}

\affiliation[g]{Deep Space Exploration Laboratory, Hefei 230088, China.}

\affiliation[h]{Dipartimento di Fisica ``Ettore Pancini'', Complesso Universitario 
di Monte S. Angelo, Universit\`a degli Studi di Napoli ``Federico II'', Via Cinthia Edificio 6, 80126 Napoli, Italy.}
 
\affiliation[i]{Departamento de Matem\'{a}ticas, Universidad Cat\'{o}lica del Norte, Avda. Angamos 0610, Casilla 1280 Antofagasta, Chile.}

\emailAdd{t.papanikolaou@ssmeridionale.it}
\emailAdd{shreya@iitism.ac.in}
\emailAdd{yifucai@ustc.edu.cn}
\emailAdd{capozziello@na.infn.it}
\emailAdd{msaridak@noa.gr}

\abstract{We present a novel model-independent generic mechanism for primordial black hole  formation within the context of non-singular matter bouncing cosmology. In particular, considering a short transition from the matter contracting phase to the Hot Big Bang expanding Universe, we find naturally enhanced curvature perturbations on very small scales which can collapse and form primordial black holes. Interestingly, the primordial black hole masses that we find can lie within the observationally unconstrained asteroid-mass window, potentially explaining the totality of dark matter. Remarkably, the enhanced curvature perturbations, collapsing to primordial black holes, can induce as well a stochastic gravitational-wave  background, being potentially detectable by future  experiments, in particular by SKA, PTAs, LISA and ET, hence  serving as a new portal to probe the  bouncing nature of the initial conditions prevailing in the early Universe.
}


\begin{document}

\maketitle
\vspace{0.6cm}
\section{Introduction}
The Hot Big Bang (HBB)~\cite{Turner:1994hv} cosmological paradigm, despite its success to describe the origin of the Universe and, more specifically, the abundances of the light elements and the origin of the isotropic cosmic microwave background (CMB), suffers from many issues, most importantly the cosmological horizon and the flatness problems. In order to address such issues, inflationary theory was introduced in early '80s~\cite{Starobinsky:1980te,Guth:1980zm,Linde:1981mu,Albrecht:1982wi,
Linde:1983gd}, being able to explain as well the origin of the large-scale structures (LSS) of the Universe.

An attractive alternative to the inflationary paradigm is the non-singular bouncing  
cosmology~\cite{Mukhanov:1991zn,Brandenberger:1993ef}, which postulates that the Universe was always contracting before the HBB era and, at some point, transitioned into the expanding Universe we are observing. This cosmological scenario is free of the initial singularity problem present in inflationary cosmology~\cite{Borde:1996pt}, solving as well the  flatness and horizon 
problems of the standard HBB theory [see Ref.~\cite{Novello:2008ra} for a review on the topic] and giving rise to scale-invariant curvature power spectra on large scales \cite{Lilley:2015ksa, Battefeld:2014uga, 
Peter:2008qz}, hence being compatible with CMB observations~\cite{Cai:2014bea,Cai:2014xxa}.

In order to accommodate a non-singular bouncing phase,  one needs to introduce an effective violation of the null energy condition for a short period of time. Consequently, modified gravity theories \cite{CANTATA:2021ktz,Nojiri:2006ri,Capozziello:2011et}  
provide us with an ideal landscape where one can realise easily a bouncing cosmological behaviour. Indicatively, let us mention that
bouncing cosmological solutions have been constructed within Pre-Big-Bang~\cite{Veneziano:1991ek} 
and Ekpyrotic~\cite{Khoury:2001wf,Khoury:2001bz} setups, higher order gravitational 
theories~\cite{Biswas:2005qr,Nojiri:2013ru, Miranda:2022wvr}, $f(R)$ 
gravity~\cite{Bamba:2013fha,Nojiri:2014zqa}, $f(T)$ 
gravity~\cite{Cai:2011tc}, $f(Q)$ gravity \cite{Bajardi:2020fxh}, non-relativistic gravity~\cite{Cai:2009in,Saridakis:2009bv}, massive 
gravity~\cite{Cai:2012ag}, braneworld scenarios~\cite{Shtanov:2002mb,Saridakis:2007cf}, loop quantum gravity~\cite{Wilson-Ewing:2012lmx,Barca:2021qdn} as well within DHOST and cyclic~\cite{Lehners:2008vx,Banerjee:2016hom,Saridakis:2018fth} cosmological models~\cite{Ilyas:2020qja,Ilyas:2020zcb,Zhu:2021whu}.

On the other hand, primordial black holes (PBHs), introduced back in `70s~\cite{1967SvA....10..602Z, Carr:1974nx,1975ApJ...201....1C,1979A&A....80..104N} can form in the early Universe before star formation out of the collapse of enhanced cosmological perturbations on small scales [See here~\cite{Khlopov:2008qy,Carr:2020gox} for nice reviews on the topic] compared to the ones probed by CMB and LSS scales. Remarkably, PBHs have rekindled the interest of the scientific community since, among others, they can account for a part or the totality of the dark matter density~\cite{Chapline:1975ojl,Belotsky:2014kca} and explain the LSS formation through Poisson fluctuations \cite{Meszaros:1975ef,Afshordi:2003zb}, providing as well the seeds for the supermassive black holes residing in the galactic centres~\cite{1984MNRAS.206..315C,Bean:2002kx}. Interestingly enough, PBHs are associated as well with numerous gravitational-wave (GW) signals originated from both binary merging events and stochastic cosmological sources~\cite{Sasaki:2018dmp,LISACosmologyWorkingGroup:2023njw}. Observational evidence for their existence can be found in \cite{Carr:2023tpt}.

An interesting way to probe non-singular bouncing cosmological scenarios is thus by exploring their interplay with PBHs~\footnote{PBHs have been extensively studied as well within the context of many physical setups alternative to the standardly studied ultra-slow-roll (USR)/inflection point inflation such as phase transitions~\cite{Jedamzik:1996mr,Niemeyer:1997mt}, false vacuum trapping~\cite{Caravano:2024tlp}, early matter era~\cite{Khlopov:1980mg,Polnarev:1986bi,Green:1997pr}, scalar field instabilities~\cite{Khlopov:1985fch}, modified/quantum gravity~\cite{Barrow:1996jk,Kawai:2021edk,Papanikolaou:2023crz} and topological defects~\cite{Polnarev:1988dh}.}. Up to now, some first attempts to bridge PBHs with bounce realizations have been performed, in particular by studying PBH formation during a matter contracting phase both analytically~\cite{Carr:2011hv,Quintin:2016qro,Chen:2016kjx,Clifton:2017hvg} and numerically~\cite{Chen:2022usd}. PBH formation was studied as well during the HBB expanding era but only within the framework of $f(R)$ gravity \cite{Banerjee:2022xft}. In this paper, we find within non-singular matter bounce cosmological scenarios a natural model-independent mechanism for PBH formation during the HBB expanding era. Furthermore,  we study
the induced GWs due to second order gravitational interactions
associated to PBH formation/production [see \cite{Domenech:2021ztg} for a review on the topic].

The paper is organised as follows: In \Sec{sec:bounce}, we introduce a model-independent parametrization of the cosmic expansion within non-singular matter bouncing frameworks, studying additionally the background and perturbation dynamics and deriving ultimately the curvature power spectrum responsible for PBH formation during the HBB expanding era. Then, in \Sec{sec:PBH}, we review the basics of PBH formation within peak theory, computing at the end the PBH abundances at our present epoch and their contribution to dark matter. Moreover, in \Sec{sec:SIGW}, we investigate the second order GWs induced by the enhanced cosmological perturbations collapsing to PBHs, checking as well their detectability with current and future GW experiments. Finally, \Sec{sec:Conclusions} is devoted to conclusions.

\section{Non-singular bouncing cosmology}\label{sec:bounce}
\subsection{Background dynamics}
Let us  consider a non-singular bouncing model which starts with a contracting matter-dominated phase, experiencing then a non-singular bouncing phase, entering finally into the HBB radiation-dominated expanding phase. Let us assume that the bouncing phase lasts from $t_{-}$ to $t_{+}$ with $t=0$ being the cosmic time at the bouncing point where the Hubble parameter vanishes, i.e. $H = 0$. For $t<<t_{-}$, the Universe is in the matter contracting phase, while for $t>>t_+$, one meets the expanding era. 

Focusing   on the background dynamics, under the aforementioned assumptions one can show that the scale factor can be approximately parameterized for each phase as~\cite{Cai:2012va,Cai:2013kja}.

\textit{(i) Contracting Phase ($t<t_{-}$):}
\begin{equation}\label{eq:a_contracting}
 a(t)=a_{-} \left(\frac{t-\tilde{t}_{-}}{t_{-}-\tilde{t}_{-}}\right)^{2/3}~,
\end{equation}
where $a_{-}$ is the scale factor at time $t_{-}$. If $H_-$ is the Hubble parameter at $t_{-}$, then one finds that $t_{-}-\tilde{t}_{-}=\frac{2}{3H_{-}}$.  One should note here that $\tilde{t}_-$ in \Eq{eq:a_contracting} is a negative integration constant which is introduced to match the Hubble parameter continuously at the time $t_-$. During the contracting phase, $t$ is negative but since $\tilde{t}_-<t<t_-$, the ratio $\frac{t-\tilde{t}_-}{t_- -\tilde{t}_-}$ is always positive, leading to a decreasing positive scale factor.

\textit{(ii) Bouncing Phase ($t_-\leq t\leq t_+$):}
\begin{equation}
a(t)=a_\text b e^{\frac{\Upsilon t^2}{2}}~,
\label{abounce}
\end{equation}
with $a_\text b$   the scale factor at the bouncing point ($t=0$) and $\Upsilon$   a model parameter depending on the underlying gravity theory driving the bounce. Matching the scale factors at $t_-$ one obtains $a_{-}=a_\text b \exp [\Upsilon t_{-}^2/{2}]$, while the corresponding Hubble parameter can be recast as
\begin{equation}
\label{bouncingH}
 H(t)=\Upsilon t~.
\end{equation}

\textit{(iii) Hot Big Bang Expanding Phase ($t> t_+$):}
\begin{equation}\label{eq:a_HBB_era_cosmic_time}
 a(t)=a_{+} \left(\frac{t-\tilde{t}_{+}}{t_{+}-\tilde{t}_{+}}\right)^{1/2}~,
\end{equation}
where $t_{+}=H_{+}/\Upsilon$ and $t_{+}-\tilde{t}_{+}=\frac{1}{2H_{+}}$. Imposing again the continuity of the scale factor at $t=t_{+}$, one acquires $a_{+}=a_\text b e^{\frac{\Upsilon t_{+}^2}{2}}$. 

 The perturbation modes exit the Hubble radius in the contracting phase, re-enter the Hubble radius in and around the bouncing phase and, after exiting the Hubble radius re-enter once again in the expanding phase. Without considering any particular model, in the next section, we  study the evolution of the perturbation modes in Fourier space through each of these phases separately in a model independent way.

\subsection{Perturbation dynamics}

Let us proceed now by considering  the perturbation behaviour. In order to make the calculation simpler, we  will work  in terms of the Mukhanov-Sasaki (MS) variable $v_k$, being related to the comoving curvature perturbation $\mathcal{R}_k$ as $v_k=z \mathcal R_k$ with $z=\frac{a \sqrt{\rho+p}}{c_s H\Mp}$. Here $c_\mathrm{s}$ stands for the curvature perturbation sound speed and $\Mp$ for the reduced Planck mass, while $\rho$ and $p$ are the energy and pressure densities, respectively.
\begin{itemize}
    \item \textbf{Evolution of the curvature perturbation during the matter contracting phase
}

Working in terms of the conformal time $\eta$ defined as $\mathrm{d}\eta \equiv \mathrm{d}t/a$, the Fourier modes of the MS variable $v_k$ will evolve according to the following equation of motion:
\begin{equation}
\label{EoM}
v_k''+\left(c_\mathrm{s,m}^2 k^2-\frac{z''}{z}\right)v_k=0~,
\end{equation}
where $c_\mathrm{s,m}$ is the sound speed during the matter contracting phase and prime denotes differentiation with respect to the conformal time. For a matter-dominated era one has that $p=0$, while the scale factor   scales as $a\propto \eta^2$. Imposing then the Bunch-Davies vacuum as our initial condition, one can write the MS variable deep in the sub-horizon regime as
\beq
v_k(k\gg aH) \simeq \frac{e^{-ik\eta}}{\sqrt{2k}},
\eeq
obtaining  at the end  $v_k$ during the matter contracting phase, reading as
\begin{equation}
\label{eq:MS_solution_contracting}
v^\mathrm{m}_k=\frac{\sqrt{\pi(-\eta)}}{2} H_{3/2}^{(1)}[c_{s,m} k (-\eta)]~,
\end{equation}
where $H_{3/2}^{(1)}$ is the $\frac{3}{2}$-order Hankel function of the first kind. Finally, the curvature power spectrum defined as $\mathcal{P}_\mathcal{R}(k)\equiv \frac{k^3}{2\pi^2}|\mathcal{R}_k|^2$ will be written as 
\beq\label{eq:P_R_contracting}
\mathcal{P}_\mathcal{R}(k) = \frac{k^3}{2\pi^2}\left|\frac{v_k}{z}\right|^2 = \frac{c^2_\mathrm{s,m}k^3(-\eta)}{24\pi\Mp^2a^2}\left|H_{3/2}^{(1)}[c_\mathrm{s,m} k (-\eta)]\right|^2.
\eeq
On large scales, i.e. $c_\mathrm{s,m}k\ll |aH|$, one obtains an almost scale invariant but time-dependent curvature power spectrum reading as $\mathcal{P}_\mathcal{R}(k) \simeq \frac{a^3_{-}H^2_{-}}{48\pi^2c_\mathrm{s,m}\Mp^2a^3}$, a result which is totally different with the superhorizon evolution in an expanding Universe being characterised by a time-independent curvature power spectrum. In our case, in contrast to an expanding phase, $\mathcal{P}_\mathcal{R}(k)$ actually grows with time, since  in a contracting phase  $a$ is decreasing with time. On the other hand, for small scales,  i.e. $c_\mathrm{s,m}k\gg |aH|$, one can show that $\mathcal{P}_\mathcal{R}(k)\simeq \frac{a^3_{-}H^2_{-}}{12\pi^2c_\mathrm{s,m}\Mp^2a^3}\left(\frac{c_\mathrm{s,m}k}{aH}\right)^2 $.

\item \textbf{Evolution of the curvature perturbation during the bouncing phase}

In the following, we restrict our analysis to a short duration bouncing phase, hence we keep all the quantities up to first order in terms of $(\eta-\eta_b)$, where $\eta_\mathrm{b}$ is the conformal time at the bouncing point, which we normalise to $0$.

From \Eq{abounce}  the expression for the scale factor in terms of $\eta$ is
\begin{equation}
    a(\eta)= a_b e^{\mathrm{InverseErf}[a_b \sqrt{2/\pi} (\eta - \eta_b) \sqrt{\Upsilon}]^2},
\end{equation}
where $a_b$ is the scale factor at the time of bounce. Keeping only terms of the order $(\eta-\eta_b)$, $z''/z$ for the present case simplifies to $a_b^2 \Upsilon$. Normalising then $a_b=1$, we once again solve the MS equation \eqref{EoM}, setting the boundary condition $v_k(\eta_-)$ for the MS variable at $\eta=\eta_-=H_-/\Upsilon$, where $v_k(\eta_-)$ is set equal to \Eq{eq:MS_solution_contracting}. The MS equation reads now as
\begin{equation}
\label{EoM2}
v_k''+(c_\mathrm{s,b}^2 k^2-\Upsilon)v_k=0~,
\end{equation}
whose solution reads as
\beq
\begin{split}
v^\mathrm{b}_k & (\eta)   =  \frac{\sqrt{\pi}}{2 H_- \sqrt{c^2_\mathrm{s,b} k^2 - \Upsilon}}
\left\{-c_\mathrm{s,m}k\frac{H^{3/2}_-}{\Upsilon^{1/2}}
H^{(1)}_{1/2}\left(\frac{c_\mathrm{s,m} H_- k}{\Upsilon}\right) \sinh\left(\frac{
\sqrt{c^2_\mathrm{s,b} k^2 - \Upsilon} (H_- + \eta \Upsilon)}{\Upsilon}\right) 
\right.
\\ & + H^{(1)}_{3/2}\left(\frac{c_\mathrm{s,m} H_- k}{\Upsilon}\right) 
\left[H_- \sqrt{c^2_\mathrm{s,b} k^2 - \Upsilon} \sqrt{\frac{c_\mathrm{s,m} H_- k}{\Upsilon}}\cosh\left(\frac{\sqrt{c^2_\mathrm{s,b} k^2 - \Upsilon} (H_- + \eta \Upsilon)}{\Upsilon}\right)\right.   \\ 
&\left. \left.   +\sqrt{\frac{H_-}{\Upsilon}} \Upsilon \sinh\left(\frac{\sqrt{c^2_\mathrm{s,b} k^2 - \Upsilon} (H_- + \eta \Upsilon)}{\Upsilon}\right)  \right] \right\}
\label{eq:MS_solution_bounce}
\end{split}
\eeq

For large scales $c_\mathrm{s,b}k\ll aH$, one finds that 
\beq\label{eq:P_R_bounce_k_small}
\begin{split}
\mathcal{P}_\mathcal{R}(k\ll aH/c_\mathrm{s,b}) & \simeq - \frac{c_\mathrm{s,b}}{\pi^2 c^3_\mathrm{s,m}}\frac{\Upsilon^3}{H^4_-}\frac{1}{(2+\eta^2\Upsilon)^2}\Biggl[ c_\mathrm{s,b}\sin^2\left(\frac{H_-+\eta\Upsilon}
{\sqrt{\Upsilon}}\right) \\ &  + 2 H_- \sqrt{\frac{c_\mathrm{s,m}c_\mathrm{s,b}}{\Upsilon}}  \sqrt{\frac{c_\mathrm{s,b}k}{aH}} \sqrt{\frac{2\eta\Upsilon}{2+\eta^2\Upsilon}} \sin\left(\frac{H_-+\eta\Upsilon}
{\sqrt{\Upsilon}}\right)\cos\left(\frac{H_-+\eta\Upsilon}
{\sqrt{\Upsilon}}\right) \\ & 
+ \left(\frac{c_\mathrm{s,b}k}{aH}\right)\frac{c_\mathrm{s,m}H^2_-}{\Upsilon}\frac{2\eta\Upsilon}{2+\eta^2\Upsilon}\cos^2\left(\frac{H_-+\eta\Upsilon}
{\sqrt{\Upsilon}}\right) \Biggr].
\end{split}
\eeq

\item \textbf{Evolution of the perturbation during the HBB expanding phase}

In the HBB expanding era, one can rewrite the scale factor \eqref{eq:a_HBB_era_cosmic_time} in terms of the conformal time as
\begin{equation}\label{eq:a_HBB_conformal_time}
   a(\eta)=\frac{H_+^2 + 2 \Upsilon}{4 \Upsilon^3}(H_+^4 + 2 H_+^2 \Upsilon - 2 \Upsilon^2 - 
   H_+ \Upsilon (H_+^2 + 2 \Upsilon) \eta). 
\end{equation}
Regarding $z(\eta)$ during the HBB expanding era, given the fact that we are in a RD era, namely $w=1/3$, we deduce that $z(\eta)$ becomes equal to $2 a(\eta)$~\footnote{During the HBB era, the underlying theory of gravity is assumed to be   General Relativity and therefore the perturbation sound speed $c_\mathrm{s}$ is equal to unity.}. Thus, accounting for \Eq{eq:a_HBB_conformal_time} one finds that $z^{\prime\prime}/z$ is $0$. Consequently, the corresponding MS equation  takes the form  of a harmonic oscillator, namely
\begin{equation}
\label{EoM3}
v_k''+k^2v_k=0~.
\end{equation}
Hence, imposing  the initial conditions at $\eta=\eta_+=H_+/\Upsilon$ as $v_k(\eta_+) = v^\mathrm{b}_k(\eta_+)$, where $v^\mathrm{b}_k$ is given by \Eq{eq:MS_solution_bounce},  to ensure the continuity of the MS variable, we acquire that the Fourier mode of the MS variable during the HBB expanding era can be recast as
\beq\label{eq:v_RD}
\begin{split}
v^\mathrm{RD}_k(\eta)  = & -\frac{c_\mathrm{s,m}\Upsilon^{3/2}}{\sqrt{2 c^2_\mathrm{s,b} k^2 - 2 \Upsilon}(c_\mathrm{s,m} H_- k)^{5/2}}
e^{\frac{i c_\mathrm{s,m} H_- k}{
\Upsilon}}\\ &  \times 
\left\{\sqrt{c^2_\mathrm{s,b} k^2-\Upsilon} \cosh\left(\frac{(H_- + H_+) \sqrt{c^2_\mathrm{s,b} k^2 -\Upsilon}}{
\Upsilon}\right) \right.
\\ & 
\times \Biggl\{\frac{(c_\mathrm{s,m} H_- k + i \Upsilon) \left(\frac{c_\mathrm{s,m} H_- k}{\Upsilon}\right)^{3/2}
\cos\left[k \left(\eta - \frac{H_+}{\Upsilon}\right)\right]}{c_\mathrm{s,m}} \\ & + \frac{\sqrt{\frac{H_-}{\Upsilon}} (-i c_\mathrm{s,m}^2 H_-^2 k^2 + c_\mathrm{s,m} H_- k \Upsilon + i \Upsilon^2) \sin\left[k \left(\eta - \frac{H_+}{\Upsilon}\right)\right]}{
\Upsilon}\Biggr\} \\ & + \biggl\{k \sqrt{\frac{H_-}{Y}} (-i c_\mathrm{s,m}^2 H_-^2 k^2 + c_\mathrm{s,m} H_- k \Upsilon + i \Upsilon^2) \cos\left[k \left(\eta - \frac{H_+}{\Upsilon}\right)\right]  \\ &+ 
H_- (c^2_\mathrm{s,b} k^2 - \Upsilon) (c_\mathrm{s,m} H_- k + i \Upsilon) \sqrt{\frac{c_\mathrm{s,m} H_- k}{\Upsilon}}
  \sin\left[k \left(\eta - \frac{H_+}{\Upsilon}\right)\right] \biggr\} \\ & \left.\times\frac{\sinh\left[\frac{(H_- + H_+) \sqrt{
c^2_\mathrm{s,b} k^2 - \Upsilon}}{\Upsilon}\right]}{\Upsilon}\right\}.
\end{split}
\eeq
\end{itemize}

\subsection{The curvature power spectrum during the Hot Bing Bang era}

\begin{figure*}[t!]
\begin{center}
\includegraphics[width=0.796\textwidth]{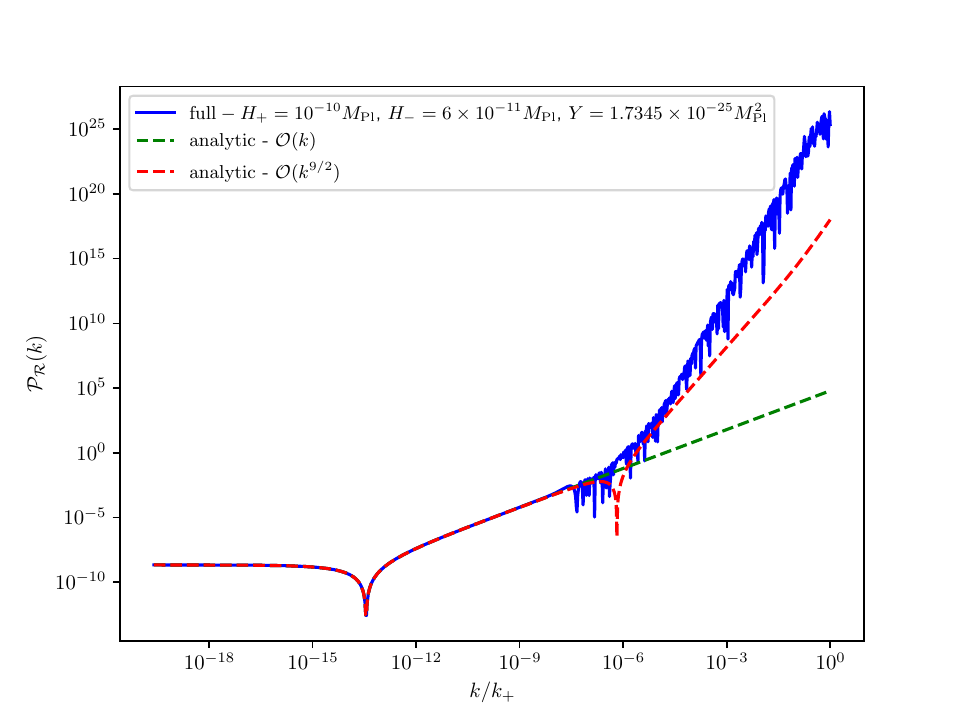}
\caption{{\it{The solid blue curve corresponds to the full curvature power spectrum, for $H_{+} = 10^{-10}\Mp$, $H_{-} = 6\times 10^{-11}\Mp$ and $\Upsilon = 1.7345\times 10^{-10}\Mp^2$. The dashed green curve corresponds to the analytic approximation for $\mathcal{P}_{\mathcal{R}}(k)$ up to linear order in $k$, while the dashed red curve  depicts the  analytic approximation for $\mathcal{P}_{\mathcal{R}}(k)$ up $\mathcal{O}(k^{9/2})$.}}}
\label{fig:P_R_full_vs_analytic}
\end{center}
\end{figure*}

The curvature power spectrum, responsible for PBH formation during the HBB era, will be the one at horizon crossing time, being considered as the typical PBH formation time, at least for nearly monochromatic PBH mass distributions. Accounting thus for the fact that the comoving curvature perturbation at superhorizon scales, during the HBB expanding era, is conserved, we can derive the curvature power spectrum at PBH formation time by setting $k=aH$. Expanding then $\mathcal{P}_\mathcal{R}(k)$ during the expansion era with respect to $k$, we extract the following analytical formula for the $\mathcal{P}_\mathcal{R}(k)$ at PBH formation time
\begin{eqnarray}\label{eq:P_R_anal}
\mathcal{P}_{\mathcal{R}}(k) &\simeq & \frac{0.7 \Upsilon^8 \cos^2{A}^2}{c_\mathrm{s,m}^3 H_-^4 H_+^2 \pi^2 (H_+^2 + 2 \Upsilon)^4} \nonumber \\ &&-\frac{1.4 B^2 \sqrt{c_\mathrm{s,m}} \Upsilon^{17/2}
     \cos{A} \sin{A} \sqrt{k}}{c_{s,m}^3 H_-^4 H_+^2 \pi^2 (H_+^2 + 2 \Upsilon)^4} \nonumber \\ && +\frac{\Upsilon^5 \biggl[0.7 c_\mathrm{s,m} H_-^3 \sin{A}^2 + 
     0.9 B^2
       \Upsilon \cos{A}\left(-\frac{2  \Upsilon^2  \cos{A}}{
         H_+ (H_+^2 + 2 \Upsilon)} + 
         \sqrt{\Upsilon} \sin{A}\right)\biggr] k}{4 c_\mathrm{s,m}^3 H_-^5 H_+^2 \pi^2 \left(1 + \frac{H_+^2}{
     2 \Upsilon}\right)^2 (H_+^2 + 2 \Upsilon)^2}, \nonumber \\  
\end{eqnarray}
where $A=(H_- + H_+)/\sqrt{\Upsilon}$ and $B=\sqrt{H_-/\Upsilon}$. 
As one may see from \Eq{eq:P_R_anal}, the first term provides the scale invariant contribution favored by CMB observations on large scales, while the second and the third terms are responsible for the enhancement of $\mathcal{P}_{\mathcal{R}}(k)$ on small scales, leading to PBH formation. As one may see from \Fig{fig:P_R_full_vs_analytic}, the analytic approximate expression for $\mathcal{P}_{\mathcal{R}}(k)$ (green dashed curve) can reproduce quite efficiently the full result (blue curve) at least within the linear regime where $\mathcal{P}_{\mathcal{R}}(k)<1$. As one proceeds to the non-linear regime, namely on very small scales, one needs to expand $\mathcal{P}_{\mathcal{R}}(k)$ to higher orders in $k$ in order to incorporate the non-linear behavior. In \Fig{fig:P_R_full_vs_analytic}, with the red dashed curve we depict the approximate formula for $\mathcal{P}_{\mathcal{R}}(k)$ up to $\mathcal{O}(k^{9/2})$. 

Furthermore, let us discuss the scaling behaviour of $\mathcal{P}_{\mathcal{R}}(k)$ as we go to smaller scales, namely higher values of $k$. In particular, in order to understand the behaviour of $\mathcal{P}_{\mathcal{R}}(k)$ at horizon crossing time during the HBB expanding phase, we should take into account the fact that the curvature perturbation is conserved on super-horizon scales in an expanding Universe. Hence, the behaviour of $\mathcal{P}_{\mathcal{R}}(k)$ at horizon crossing time during the HBB expanding phase will be dictated by its behaviour on super-horizon scales during the bouncing phase. 

Interestingly enough, as one may infer from \Eq{eq:P_R_bounce_k_small} for very large scales, i.e. $c_\mathrm{s,b}k\gg aH$, this equation gives a scale-independent $\mathcal{P}_{\mathcal{R}}(k)$, as the one shown in \Fig{fig:P_R_full_vs_analytic} and extracted in the approximate formula \eqref{eq:P_R_anal} for the HBB expanding phase. Then, as we go to smaller scales, however remaining always  within the super-horizon regime, \Eq{eq:P_R_bounce_k_small} starts to be dominated by the term linear in $k$, being in agreement with the linear growth of $\mathcal{P}_{\mathcal{R}}(k)$ shown in \Eq{eq:P_R_anal}. If now one goes to even smaller scales, they will depart from the linear growth scaling of $\mathcal{P}_{\mathcal{R}}(k)$, starting to exhibit strong oscillatory features. 

The difference from the linear scaling behaviour can be revealed if we expand $\mathcal{P}_{\mathcal{R}}(k)$ beyond linear order, while  the oscillatory behaviour comes from the fact that, as we go close to $k=k_+$, with $k_+$ being the mode crossing the horizon at the onset of the HBB expanding phase, the term $e^{\frac{i c_\mathrm{s,m} H_- k}{
\Upsilon}} \Bigg\{\sin\left[k \left(\eta - \frac{H_+}{\Upsilon}\right)\right]\quad\mathrm{or}\quad\cos\left[k \left(\eta - \frac{H_+}{\Upsilon}\right)\right]\Biggr\}$ in \Eq{eq:v_RD} will enter to a resonant regime yielding strong oscillations. This can be interpreted physically by the fact that $k_+$ is the smallest scale of our scenario. All modes with $k>k_+$ are always sub-horizon in all   three regimes, namely contracting, bouncing and expanding phases, being characterised by strong oscillatory behaviours. Therefore, modes  which are slightly larger than $k^{-1}_+$ will pass a very short period in the super-horizon regime, being most of the time sub-horizon during the bouncing phase.

At this point  it is important to emphasize that the growth of curvature perturbations on small scales is a generic feature of any non-singular matter bouncing cosmological setup. 
This is physically justified due to the growth of the curvature perturbations on super-horizon scales during the matter contracting phase, independently of the parametrisation of the scale factor during the bouncing phase [See \Eq{eq:P_R_contracting}]. In particular, both the amplitude and the shape of the power spectrum of primordial curvature perturbations remain unchanged through the bounce due to a no-go theorem~\cite{Quintin:2015rta,Battarra:2014tga}, independently of the duration of the bouncing phase.
Hence, one  can acquire a generic non-``fine-tuned" mechanism of PBH formation within non-singular matter bouncing cosmology, in contrast with the ``fine-tuned'' PBH formation present in single-field ultra-slow roll inflationary setups~\cite{Cole:2023wyx}.

The fine-tuning of the order of $10^{-4}$ at the level of $\Upsilon$ (see e.g \Fig{fig:P_R_full_vs_analytic}) is due to the fact that once fixing $H_+$ and $H_-$ one should ``fine-tune" the value of $\Upsilon$ in order to obtain a scale-invariant curvature power spectrum on CMB scales, i.e. require that the first term of \Eq{eq:P_R_anal} is equal to $2.1\times 10^{-9}$ as imposed by Planck~\cite{Planck:2018vyg}, namely
\beq\label{eq:Y_fine_tuning}
\frac{0.7 \Upsilon^8 \cos^2\left(\frac{H_+ + H_-}{\sqrt{\Upsilon}}\right)^2}{c_\mathrm{s,m}^3 H_-^4 H_+^2 \pi^2 (H_+^2 + 2 \Upsilon)^4}  = 2.1\times 10^{-9}.
\eeq
\Eq{eq:Y_fine_tuning} is a complicated algebraic equation, with $\Upsilon$ appearing  inside and outside the cosine, giving rise to the fine-tuning of $\Upsilon$.

\section{Primordial black hole formation in the expanding Hot Big Bang era}\label{sec:PBH}

Having found in the previous section an enhanced curvature power spectrum on small scales favoring PBH production, let us now review  the basics of PBH formation, calculating at the end the PBH abundances within our non-singular bouncing cosmological scenario. In the following, we will consider PBH formation due to the gravitational collapse of enhanced cosmological perturbations re-crossing  the cosmological horizon during the expanding HBB radiation-dominated (RD) era. In particular, we will determine the PBH abundance within the peak theory and ultimately the fraction of dark matter in form of PBHs.

\subsection{Basic steps of primordial black hole formation}

Considering spherical symmetry on super-horizon scales, the metric describing the collapsing overdensity region can be recast as~\cite{Starobinsky:1982ee}
\beq\label{eq:metric_spherical}
\mathrm{d}s^2 = -\mathrm{d}t^2 + a^2(t)e^{2\mathcal{R}(r)}\left[\mathrm{d}r^2 + r^2\mathrm{d}\Omega^2\right],
\eeq
where $a(t)$ is the scale factor and $\mathcal{R}(r)$ is the comoving curvature perturbation being conserved on super-horizon scales in an expanding cosmological era~\cite{Wands:2000dp}.
$\mathcal{R}(r)$ is actually related to the energy density contrast in the comoving gauge as
\begin{equation}\label{eq:zeta_vs_delta:non_linear}
\begin{split}
\frac{\delta\rho}{\rho_\mathrm{b}} &\equiv \frac{\rho(r,t)-\rho_{\mathrm{b}}(t)}{\rho_{\mathrm{b}}(t)} \\ & = -\left(\frac{1}{aH}\right)^2\frac{4(1+w)}{5+3w}e^{-5\mathcal{R}(r)(r)/2}\nabla^2e^{\mathcal{R}(r)/2},
\end{split}
\end{equation}
with $H(t) = \dot{a}(t)/a(t)$ being the Hubble parameter and $w$  the equation-of-state (EoS) parameter $w\equiv p/\rho$.  In the linear regime ($\mathcal{R}\ll 1$), \Eq{eq:zeta_vs_delta:non_linear} is written as
\beq\label{eq:zeta_vs_delta:linear}
\frac{\delta\rho}{\rho_\mathrm{b}}\simeq -\frac{1}{a^2H^2}\frac{2(1+w)}{5+3w}\nabla^2\mathcal{R}(r) \Longrightarrow \delta_k =  \frac{k^2}{a^2H^2}\frac{2(1+w)}{5+3w}\mathcal{R}_k.
\eeq
Note that due to the $k^2$ damping, large scales that cannot be observed are naturally removed~\footnote{Working in terms of  comoving curvature perturbation $\mathcal{R}$, PBH abundances are significantly overestimated, since large unobservable scales are not removed when smoothing the PBH distribution~\cite{Young:2014ana}.}.

Let us emphasize here that PBH formation is a non-linear process. One should then in principle consider the full non-linear relation \eqref{eq:zeta_vs_delta:non_linear} between $\mathcal{R}$ and $\delta$. At the end, one can deduce that the smoothed energy density contrast, denoted as $\delta_\mathrm{m}$, scales with the linear energy 
density contrast $\delta_l$, given by \Eq{eq:zeta_vs_delta:linear}, as~\cite{DeLuca:2019qsy,Young:2019yug}
\beq\label{eq:delta_m_smoothed}
\delta_\mathrm{m} = \delta_l - \frac{3}{8}\delta^2_l,
\eeq
where scales smaller than the cosmological horizon scale have been smoothed out in order to account for the cloud-in-cloud issue, while larger scales are naturally removed due to the $k^2$ damping mentioned above. In particular, the smoothed $\delta_l$ is defined as
\beq\label{eq:smoothed_delta_l}
\delta^R_l \equiv \int \mathrm{d}^3\vec{x}^\prime W(\vec{x},R)\delta(\vec{x}-\vec{x}^\prime).
\eeq
In \Eq{eq:smoothed_delta_l}, we consider a Gaussian window function $W(\vec{x},R)$ whose expression in $k$ space reads as~\cite{Young:2014ana}
\beq\label{eq:Gaussian_window_function}
\tilde{W}(R,k) = e^{-k^2R^2/2},
\eeq
with $R$ being the smoothing scale, roughly equal to the comoving horizon scale $R=(aH)^{-1}$ for nearly monochromatic PBH mass distributions. Making use now of \Eq{eq:zeta_vs_delta:linear}, the smoothed variance of the energy density field can be recast as
\beq\label{eq:sigma}
\begin{split}
\sigma^2 & \equiv \langle \left(\delta^{R}_l\right)^2\rangle = \int_0^\infty\frac{\mathrm{d}k}{k}\mathcal{P}_{\delta_l}(k,R)  \\ & = \frac{4(1+w)^2}{(5+3w)^2}\int_0^\infty\frac{\mathrm{d}k}{k}(kR)^4 \tilde{W}^2(k,R) \mathcal{P}_\mathcal{R}(k),
\end{split}
\eeq
where $\mathcal{P}_{\delta_l}(k,R)$ and $\mathcal{P}_{\mathcal{R}}(k)$ stand for the reduced energy density and curvature power spectra respectively. 

Concerning the PBH mass, being of the order of the cosmological horizon mass at the time of PBH formation, its spectrum will follow a critical collapse scaling law~\cite{Niemeyer:1997mt,Niemeyer:1999ak,Musco:2008hv,Musco:2012au},  
\beq\label{eq:PBH_mass_scaling_law}
M_\mathrm{PBH} = M_\mathrm{H}\mathcal{K}(\delta-\delta_\mathrm{c})^\gamma,
\eeq
with $M_\mathrm{H}$ being the mass within the cosmological horizon at horizon crossing time. Here $\gamma \simeq 0.36$ is a critical exponent, depending on the EoS at PBH formation time,  being that of radiation. The parameter $\mathcal{K}$ depends on the EoS parameter as well as on the shape of the collapsing overdensity region. In the following, we will adopt a fiducial value for $\mathcal{K}\simeq 4$ based on numerical simulations of PBH formation during a RD era~\cite{Musco:2008hv}. 

With regards to the PBH formation threshold value, $\delta_\mathrm{c}$, the latter will depend, in general, on the shape of the collapsing curvature perturbation profile~\cite{Musco:2018rwt,Musco:2020jjb}, on the EoS parameter at the time of PBH formation~\cite{Harada:2013epa,Escriva:2020tak,Papanikolaou:2022cvo}, as well on the presence of anisotropies~\cite{Musco:2021sva} and non-sphericities~\cite{Yoo:2020lmg,Yoo:2024lhp}. In our case, we consider the standard case of spherical isotropic collapse in the HBB RD expanding era. Thus, we need to investigate the effect  of the collapsing curvature power spectrum profile shape on $\delta_\mathrm{c}$. In particular, as it can be seen from  Fig. \ref{fig:P_R_full_vs_analytic}, we have, in principle, broad curvature power spectra and, on very small scales where one enters the non-linear regime, i.e. $\mathcal{P}_\mathcal{R}(k)>1$, we observe oscillatory features as well. Therefore, in order to determine the value $\delta_\mathrm{c}$, we adopt the methodology introduced in~\cite{Musco:2020jjb}. 

\subsection{The primordial black hole abundance within peak theory}

Having smoothed  the above energy density field and accounted for the critical collapse scaling law PBH mass spectrum, we can now proceed to the calculation of the PBH mass function $\beta(M)$ working within the  context of peak theory. This  states that the density of sufficiently rare and large peaks for a random Gaussian density field in spherical symmetry is written as ~\cite{Bardeen:1985tr}
\beq\label{eq:peak_density}
\mathcal{N}(\nu) = \frac{\mu^3}{4\pi^2}\frac{\nu^3}{\sigma^3}e^{-\nu^2/2},
\eeq
where $\nu \equiv \delta/\sigma$ and $\sigma$ is the smoothed variance of the energy density field given by \Eq{eq:sigma}. The parameter $\mu$, appearing in \Eq{eq:peak_density}, is actually the first moment of the smoothed curvature power spectrum defined as
\beq
\begin{split}
\mu^2 & =\int_0^\infty\frac{\mathrm{d}k}{k}\mathcal{P}_{\delta_l}(k,R)\left(\frac{k}{aH}\right)^2 \\ & = \frac{4(1+w)^2}{(5+3w)^2}\int_0^\infty\frac{\mathrm{d}k}{k}(kR)^4 \tilde{W}^2(k,R)\mathcal{P}_\mathcal{R}(k)\left(\frac{k}{aH}\right)^2.
\end{split}
\eeq
Thus, the fraction  of the Universe at a peak of a given height $\nu$ collapsing to form a PBH, denoted here as $\beta_\nu$, reads as 
\beq
\beta_\nu = \frac{M_\mathrm{PBH}(\nu)}{M_\mathrm{H}}\mathcal{N}(\nu)\Theta(\nu - \nu_\mathrm{c}),
\eeq
and the total energy density contribution of PBHs of mass $M$ to the energy budget of the Universe, namely the PBH mass function, is 
\beq\label{eq:beta_full_non_linear}
\beta(M) = \int_{\nu_\mathrm{c}}^{\frac{4}{3\sigma}}\mathrm{d}\nu\frac{\mathcal{K}}{4\pi^2}\left(\nu\sigma - \frac{3}{8}\nu^2\sigma^2 - \delta_{\mathrm{c}}\right)^\gamma \frac{\mu^3\nu^3}{\sigma^3}e^{-\nu^2/2},
\eeq
where $\nu_\mathrm{c} = \delta_{\mathrm{c},l}/\sigma$ and $\delta_{\mathrm{c},l}=\frac{4}{3}\left(1 -
\sqrt{\frac{2-3\delta_\mathrm{c}}{2}}\right)$.

One can then extract the PBH abundance and its contribution to the dark matter abundance. Doing so, we introduce the quantity $f_\mathrm{PBH}$ defined as
\beq\label{eq:f_PBH_def}
f_\mathrm{PBH}\equiv \frac{\Omega_\mathrm{PBH,0}}{\Omega_\mathrm{DM,0}},
\eeq
where the subscript $0$ refers to our present epoch and $\Omega_\mathrm{PBH}=\rho_\mathrm{PBH}/\rho_\mathrm{crit}$, $\Omega_\mathrm{DM,0} = 0.265$. Accounting now for the fact that PBHs behave as pressureless dust one has that 
$\rho_\mathrm{PBH,0}=\rho_\mathrm{PBH,f}\left(a_\mathrm{f}/a_\mathrm{0}\right)^3\simeq \beta \rho_\mathrm{rad,f}\left(a_\mathrm{f}/a_\mathrm{0}\right)^3$ where the index  ``$\mathrm{f}$" refers to PBH formation time. At the end, considering the fact that $M\simeq M_\mathrm{H}$ and applying as well entropy conservation from PBH formation time up to our present epoch, one   straightforwardly finds that  
\beq\label{eq:f_PBH}
f_\mathrm{PBH} = \left(\frac{\beta(M)}{3.27 \times 10^{-8}}\right) \left(\frac{106.75}{g_{*,\mathrm{f}}}\right)^{1/4}\left(\frac{M}{M_\odot}\right)^{-1/2},
\eeq
where $M_\odot$ is the solar mass and where $g_{*,\mathrm{f}}$ is the effective number of relativistic degrees of freedom. For our numerical applications, we will use $g_{*,\mathrm{f}}=106.75$, being  the number of relativistic degrees of freedom of the Standard Model before the electroweak phase transition~\cite{Kolb:1990vq}.

\begin{figure*}[t!]
\begin{center}
\includegraphics[height=6cm,width=7.cm]{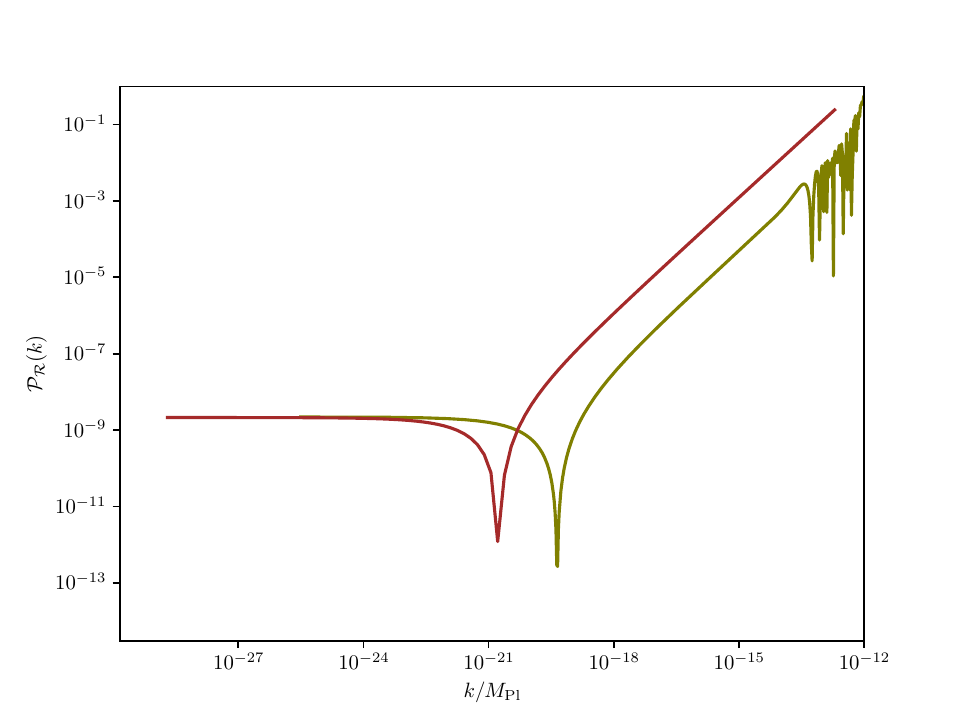}
\includegraphics[height=6cm,width=7.cm]{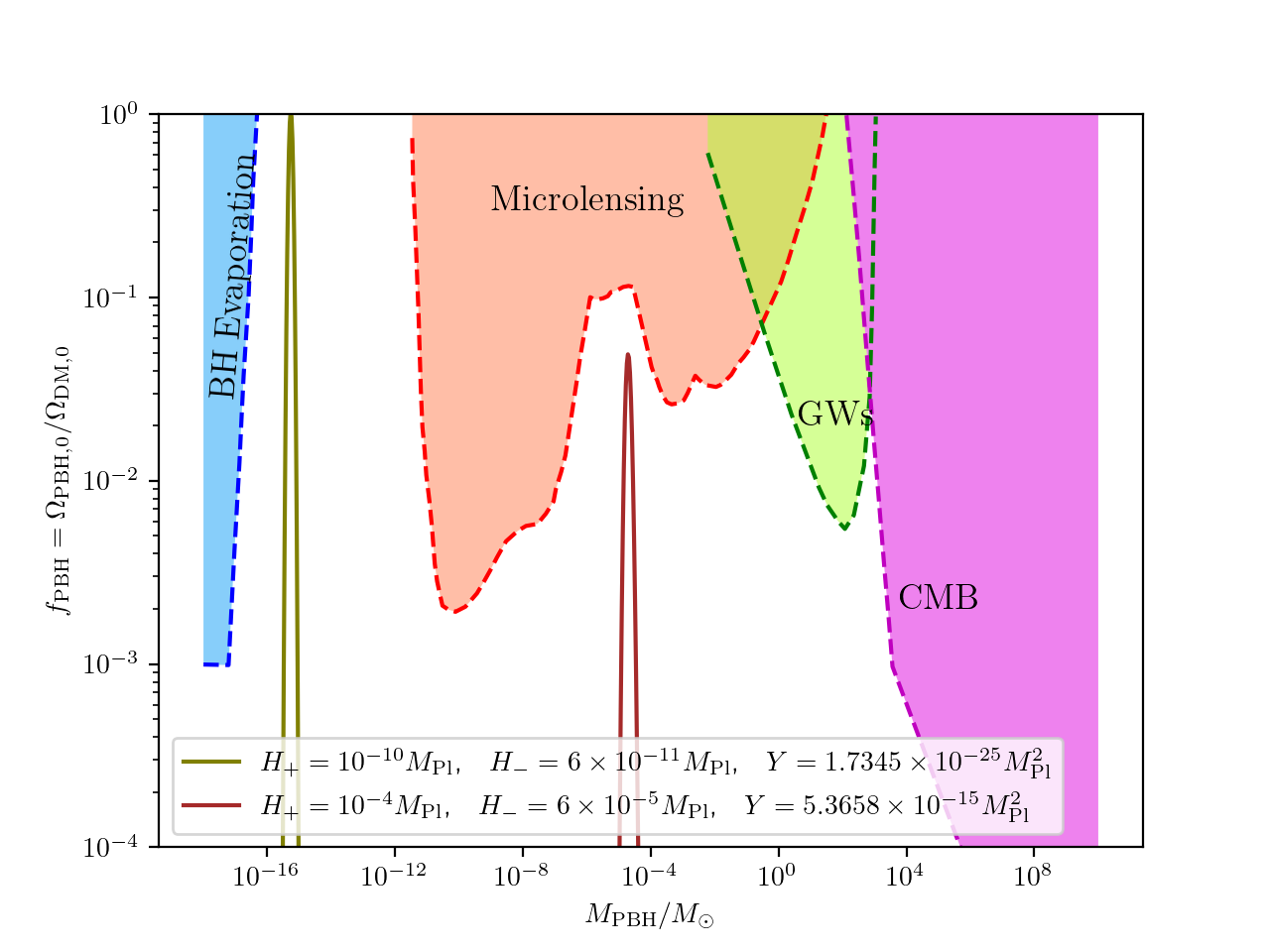}
\caption{{\it{Left Panel: The curvature power spectra for different fiducial values of the parameters $H_{+}$, $H_{-}$ and $\Upsilon$. Right Panel: The fraction of dark matter in terms of PBHs denoted as $f_\mathrm{PBH}=\Omega_\mathrm{PBH,0}/\Omega_\mathrm{DM,0}$ as a function of the PBH mass. The colored regions are excluded from evaporation (blue region), microlensing (red region), gravitational-wave (green region) and CMB (violet region) observational probes concerning the PBH abundances. The data for the constraints on $f_\mathrm{PBH}$ from the different observational probes were obtained from~\cite{Green:2020jor}.}}}
\label{fig:f_PBH}
\end{center}
\end{figure*}

In \Fig{fig:f_PBH}, we show  in the left panel  the curvature power spectra for two different sets of the theoretical parameters involved, namely $H_+$, $H_-$, and $\Upsilon$, whereas in the right panel we present,  associated to these curvature power spectra, the PBH energy density contribution to dark matter $f_\mathrm{PBH}$  as a function of the PBH mass. Additionally, we have superimposed     constraints on $f_\mathrm{PBH}$ from evaporation (blue region)~\cite{Poulin:2016anj,Clark:2016nst,Boudaud:2018hqb,DeRocco:2019fjq,Laha:2019ssq}, microlensing (red region)~\cite{Macho:2000nvd,Niikura:2017zjd,Niikura:2019kqi,Zumalacarregui:2017qqd}, GW (green region)~\cite{Kavanagh:2018ggo,Chen:2019irf} and CMB (violet region)~\cite{Serpico:2020ehh} observational probes. In~\cite{Green:2020jor} one can find a combined analysis of the aforementioned PBH abundance constraints. Regarding now the value of the PBH formation threshold computed following the procedure introduced in~\cite{Musco:2020jjb}, we found that for the case where $H_+ = 10^{-4}\Mp$, $H_- = 6\times 10^{-5}\Mp$ and $\Upsilon = 5.3658 \times 10^{-15}\Mp^2$, $\delta_\mathrm{c} = 0.575$ whereas for $H_+ = 10^{-10}\Mp$, $H_- = 6\times 10^{-11}\Mp$ and $\Upsilon = 1.7345\times 10^{-25}\Mp^2$, $\delta_\mathrm{c} = 0.582$. 

As one can see from the right panel of \Fig{fig:f_PBH}, we can produce PBHs within a wide range of masses depending on the values of $H_+$, $H_-$ and $\Upsilon$. In particular, the PBH mass will be of the order of the cosmological horizon mass at the time of PBH formation, i.e. horizon crossing time as it can be seen by \Eq{eq:PBH_mass_scaling_law}. After a straightforward calculation we can show that the typical mass of a PBH forming in the HBB expanding era will scale with $H_+$, $H_-$, $\Upsilon$ and the comoving scale $k$ as 
\beq
M_\mathrm{PBH}\simeq M_\mathrm{H} = \frac{4\pi\Mp^2}{H_\mathrm{hc}(H_+,H_-,\Upsilon,k)} = \frac{\pi\Mp^2H_+\left(H^2_+ + 2\Upsilon\right)^2}{\Upsilon^2k^2},
\eeq
where $H_\mathrm{hc}$ is the Hubble parameter at horizon crossing time.

Interestingly enough, as we can notice in the right panel of \Fig{fig:f_PBH}, we can easily produce PBHs with mass of the order of one solar mass, being the typical black hole progenitor masses for the LIGO-VIRGO-KAGRA (LVK) merging events as well PBHs within the observationally unconstrained asteroid-mass window, where PBHs can account for the totality of dark matter. 

At this point, it is important to stress that, in order to stay within the perturbative regime, we impose a non-linear cut-off scale $k_\mathrm{NL}$ depending on $H_+$, $H_-$ and $\Upsilon$ such as that $\mathcal{P}_\mathcal{R}(k_\mathrm{NL}) = 0.1$. Going beyond the non-linear regime, where cosmological perturbation theory breaks down, will require to perform high-cost $N$-body numerical simulations, which lies beyond the scope of this work. 

Let us comment here that one should account as well for the backreaction of small-scale one-loop corrections to the large-scale curvature power spectrum, which could potentially alter the curvature perturbation amplitude measured by Planck. At least within single-field inflationary models, this issue was studied~\cite{Inomata:2022yte, Kristiano:2022maq,Choudhury:2023jlt,Choudhury:2023rks,Choudhury:2023vuj,Ballesteros:2024zdp} with the more recent works claiming that it can be evaded 
\cite{Franciolini:2023lgy,Firouzjahi:2023ahg,Ballesteros:2024zdp}. It is still however an open issue what happens within alternative to inflationary setups as the one considered here. To answer this question one should perform a case-by-case study.

\section{Scalar induced gravitational waves}\label{sec:SIGW}

Having studied  the PBH formation with the context of non-singular matter bouncing cosmologies, let us proceed to the exploration of the stochastic GW background induced at second order in cosmological perturbation theory by the enhanced curvature perturbations collapsed to form PBHs~\cite{Matarrese:1992rp,Matarrese:1993zf,Matarrese:1997ay,Mollerach:2003nq}  [see~\cite{Domenech:2021ztg} for a review].

\subsection{Tensor perturbations}\label{sec:tensor_perturbations}
Working in the Newtonian gauge,~\footnote{As noted in~\cite{Hwang:2017oxa,Tomikawa:2019tvi,DeLuca:2019ufz,Yuan:2019fwv,Inomata:2019yww,Domenech:2020xin,Chang:2020tji}, there is no gauge dependence for induced scalar tensor modes during a RD era, as the one we study in this work, due to the decay of the GW source, namely the scalar perturbations, in the late-time limit.} the perturbed Friedman-Lema\^itre-Robertson-Walker (FLRW) metric can  be written as
\bea
\label{metric decomposition with tensor perturbations}
\mathrm{d}s^2 = a^2(\eta)\left\lbrace-(1+2\Phi)\mathrm{d}\eta^2  + \left[(1-2\Phi)\delta_{ij} + \frac{h_{ij}}{2}\right]\mathrm{d}x^i\mathrm{d}x^j\right\rbrace \, ,
\eea
where $\Phi$ is the first-order scalar perturbation, usually denoted as Bardeen potential, and $h_{ij}$ is second-order tensor perturbation. Going now in the Fourier space, the tensor perturbation mode $h_{ij}$ will be recast as
\beq
\label{h_ij Fourier decomposition}
h_{ij}(\eta,\boldmathsymbol{x}) = \int \frac{\mathrm{d}^3\boldmathsymbol{k}}{\left(2\pi\right)^{3/2}} \left[h^{(+)}_\boldmathsymbol{k}(\eta)e^{(+)}_{ij}(\boldmathsymbol{k}) + h^{(\times)}_\boldmathsymbol{k}(\eta)e^{(\times)}_{ij}(\boldmathsymbol{k}) \right]e^{i\boldmathsymbol{k}\cdot\boldmathsymbol{x}},
\eeq
with $e^{(+)}_{ij}$ and $e^{(-)}_{ij}$   the polarisation tensors   defined as
\begin{eqnarray}
e^{(+)}_{ij}(\boldmathsymbol{k}) \equiv \frac{1}{\sqrt{2}}\left[e_i(\boldmathsymbol{k})e_j(\boldmathsymbol{k}) - \bar{e}_i(\boldmathsymbol{k})\bar{e}_j(\boldmathsymbol{k})\right], \\ 
e^{(\times)}_{ij}(\boldmathsymbol{k}) \equiv \frac{1}{\sqrt{2}}\left[e_i(\boldmathsymbol{k})\bar{e}_j(\boldmathsymbol{k}) + \bar{e}_i(\boldmathsymbol{k})e_j(\boldmathsymbol{k})\right],
\end{eqnarray}
where $e_i(\boldmathsymbol{k})$ and $\bar{e}_i(\boldmathsymbol{k})$ are two 3D vectors which alongside with $\boldmathsymbol{k}/k$ form an orthonormal basis. Finally, the tensor modes $h_\boldmathsymbol{k}$   obey the following equation~\cite{Ananda:2006af,Baumann:2007zm}:
\beq
\label{Tensor Eq. of Motion}
h_\boldmathsymbol{k}^{s,\prime\prime} + 2\mathcal{H}h_\boldmathsymbol{k}^{s,\prime} + k^{2}h^s_\boldmathsymbol{k} = 4 S^s_\boldmathsymbol{k}\, ,
\eeq
where 
$s = (+), (\times)$ stands for the two polarisation modes of tensor perturbations in General Relativity and $S^s_\boldmathsymbol{k}$ is a source term reading as~\cite{Kohri:2018awv,Espinosa:2018eve}
\beq
\label{Source}
S^s_\boldmathsymbol{k}  =
\int\frac{\mathrm{d}^3 q}{(2\pi)^{3/2}}e^{s}(\boldmathsymbol{k},\boldmathsymbol{q})F(\boldmathsymbol{q},|\boldmathsymbol{k-q}|,\eta)\phi_\boldmathsymbol{q}\phi_\boldmathsymbol{k-q}.
\eeq
In \Eq{Source}, we have written the Fourier mode of $\Phi$  as $\Phi_k(\eta) = T_\Phi(\tilde{x})\phi_k$ with $\tilde{x}=k\eta$, where $\phi_k$ is the value of $\Phi$ at some reference time $\tilde{x}_0$ - here we consider it to be the horizon crossing time - and $T_\Phi(\tilde{x})$ is a transfer function,   defined as $T_\Phi(\tilde{x})\equiv\Phi(\tilde{x})/\Phi(\tilde{x}_0)$. For the radiation-dominated Universe  we are considering here, $T_\Phi(\tilde{x})$ takes the following form:
\begin{equation}\label{eq: transfer RD}
T(\tilde{x})= \frac{9}{\tilde{x}^2}\left[ \frac{\sin (\tilde{x}/\sqrt 3)}{\tilde{x}/\sqrt 3} -\cos(\tilde{x}/\sqrt 3) \right]. 
\end{equation}
Moreover, the function  $F(\boldmathsymbol{q},|\boldmathsymbol{k-q}|,\eta)$ can be written as
\bea
\label{F}
F(\boldmathsymbol{q},|\boldmathsymbol{k-q}|,\eta) & = 2T_\Phi(q\eta)T_\Phi\left(|\boldmathsymbol{k}-\boldmathsymbol{q}|\eta\right) 
\\  & \kern-2em + \frac{4}{3(1+w)}\left[\mathcal{H}^{-1}qT_\Phi^{\prime}(q\eta)+T_\Phi(q\eta)\right]\\  & \kern-2em \times \left[\mathcal{H}^{-1}\vert\boldmathsymbol{k}-\boldmathsymbol{q}\vert T_\Phi^{\prime}\left(|\boldmathsymbol{k}-\boldmathsymbol{q}|\eta\right)+T_\Phi\left(|\boldmathsymbol{k}-\boldmathsymbol{q}|\eta\right)\right].
\eea
Consequently, \Eq{Tensor Eq. of Motion} can be solved analytically with the use of Green function formalism, and the solution of the mode function $h_\boldmathsymbol{k}^{s}$ can be written  as~\cite{Kohri:2018awv}
\beq
\label{tensor mode function}
h^s_\boldmathsymbol{k} (\eta)  =\frac{4}{a(\eta)} \int^{\eta}_{\eta_\mathrm{d}}\mathrm{d}\bar{\eta}\,  G^s_\boldmathsymbol{k}(\eta,\bar{\eta})a(\bar{\eta})S^s_\boldmathsymbol{k}(\bar{\eta}),
\eeq
with the Green function $G^s_{\bm{k}}(\eta,\bar{\eta})$ derived from the homogeneous equation 
\beq
\label{Green function equation}
G_\boldmathsymbol{k}^{s,\prime\prime}(\eta,\bar{\eta})  + \left( k^{2} -\frac{a^{\prime\prime}}{a}\right)G^s_\boldmathsymbol{k}(\eta,\bar{\eta}) = \delta\left(\eta-\bar{\eta}\right),
\eeq
under the boundary conditions $\lim_{\eta\to \bar{\eta}}G^s_\boldmathsymbol{k}(\eta,\bar{\eta}) = 0$ and $ \lim_{\eta\to \bar{\eta}}G^{s,\prime}_\boldmathsymbol{k}(\eta,\bar{\eta})=1$.

\subsection{The scalar induced gravitational-wave signal}

Focusing now on the sub-horizon regime, where we can use the flat spacetime approximation, since  on small scales  one does not feel the curvature of space-time, we can show that the energy density of the gravitational waves can be written as~\cite{Isaacson:1968zza,Maggiore:1999vm}
\bea
\label{rho_GW effective}
 \rhoGW (\eta,\boldmathsymbol{x}) & =  \frac{\Mp^2}{32 a^2}\, \overline{\left(\partial_\eta h_\mathrm{\alpha\beta}\partial_\eta h^\mathrm{\alpha\beta} +  \partial_{i} h_\mathrm{\alpha\beta}\partial^{i}h^\mathrm{\alpha\beta} \right)},
\eea
being the sum of a gradient and a kinetic term, which, in the case of a free GW, are equipartitioned. 

In the RD era, due to diffusion damping~\cite{1980lssu.book.....P,1968ApJ...151..459S}, the scalar perturbations are decaying very fast, hence decoupling quickly from the tensor perturbations soon after horizon crossing. Thus, accounting only for sub-horizon modes and neglecting the friction term in (\ref{Tensor Eq. of Motion}), which is now suppressed, \Eq{Tensor Eq. of Motion} becomes a free-wave equation and the effective GW energy density will be given by
\beq\label{rho_GW_effective_averaged}
\begin{split}
 \left\langle \rhoGW (\eta,\boldmathsymbol{x}) \right\rangle  & \simeq 2 \sum_{s=+,\times}\frac{\Mp^2}{32a^2}\overline{\left\langle\left(\nabla h^{s}_\mathrm{\alpha\beta}\right)^2\right \rangle }
 \\ & =   \frac{\Mp^2}{16a^2 \left(2\pi\right)^3} \sum_{s=+,\times} \int\mathrm{d}^3\boldmathsymbol{k}_1 \int\mathrm{d}^3\boldmathsymbol{k}_2\,  k_1 k_2  \\ & \times \overline{  \left\langle h^{s}_{\boldmathsymbol{k}_1}(\eta)h^{s,*}_{\boldmathsymbol{k}_2}(\eta)\right\rangle} e^{i(\boldmathsymbol{k}_1-\boldmathsymbol{k}_2)\cdot \boldmathsymbol{x}}\, ,
 \end{split}
\eeq
where the bar stands for averaging over the sub-horizon oscillations of $h_\boldmathsymbol{k}$ and $\langle ... \rangle$ denotes an ensemble average. The factor  $2$ in the first line of \Eq{rho_GW_effective_averaged} appears due to the equipartition of the gradient and  the kinetic energy density terms in \Eq{rho_GW effective} in the case of a free GW.

Defining now $\OmegaGW(\eta,k)$ through the relation
\bea
\label{eq:OmegaGW:def}
\left\langle  \rhoGW (\eta,\boldmathsymbol{x}) \right\rangle \equiv \rho_\mathrm{tot} \int \OmegaGW(\eta,k)  \dd\ln k,
\eea
where $\rho_\mathrm{tot}$ is the total energy density of the Universe, we   can calculate $\OmegaGW(\eta,k)$ by computing $\left\langle  \rhoGW (\eta,\boldmathsymbol{x}) \right\rangle$. Equivalently, given \Eq{rho_GW_effective_averaged}, one obtains $\OmegaGW(\eta,k)$ by computing the two-point correlation function of the tensor field  $\langle h^r_{\boldmathsymbol{k}_1}(\eta)h^{s,*}_{\boldmathsymbol{k}_2}(\eta)\rangle$.


In \Eq{rho_GW_effective_averaged},   inside the double integral one can see the appearance of the equal time correlation function for  tensor modes, which, basically, provides     the tensor power spectrum $\mathcal{P}_{h}(\eta,k)$ through the following expression:
\bea
\label{tesnor power spectrum definition}
\langle h^r_{\boldmathsymbol{k}_1}(\eta)h^{s,*}_{\boldmathsymbol{k}_2}(\eta)\rangle \equiv \delta^{(3)}(\boldmathsymbol{k}_1 - \boldmathsymbol{k}_2) \delta^{rs} \frac{2\pi^2}{k^3_1}\mathcal{P}^{(s)}_{h}(\eta,k_1),
\eea
where again $s=(\times)$ or $(+)$. 

After very long but straightforward algebraic manipulations and considering that on the super-horizon regime $\Phi=2\mathcal{R}/3$~\cite{Mukhanov:1990me}, the tensor power spectrum $\mathcal{P}_{h}(\eta,k)$ reads as [see ~\cite{Kohri:2018awv,Espinosa:2018eve} for more details] 
\begin{equation}
\label{Tensor Power Spectrum}
\mathcal{P}^{(s)}_h(\eta,k) = 4\int_{0}^{\infty} \mathrm{d}v\int_{|1-v|}^{1+v}\mathrm{d}u   \left[ \frac{4v^2 - (1+v^2-u^2)^2}{4uv}\right]^{2} \times I^2(u,v,x)\mathcal{P}_\mathcal{R}(kv)\mathcal{P}_\mathcal{R}(ku)\,,
\end{equation}
with $I(u,v,x)$   a kernel function containing information on the thermal state of the Universe during the era of GW production, defined as 
\bea
\label{I function}
I(u,v,x) \equiv \frac{2}{3}\int_{x_\mathrm{d}}^{x} \mathrm{d}\bar{x}\, \frac{a(\bar{x})}{a(x)}\, k\, G_{k}(x,\bar{x}) F_k(u,v,\bar{x}).
\eea

Using \Eq{eq:OmegaGW:def} we can write the GW spectral density as the GW energy density per logarithmic comoving scale. Combining then \Eq{Tensor Power Spectrum} and \Eq{tesnor power spectrum definition}, and inserting \Eq{tesnor power spectrum definition} into \Eq{rho_GW_effective_averaged}, we acquire  
\beq\label{Omega_GW}
\Omega_\mathrm{GW}(\eta,k)\equiv \frac{1}{\bar{\rho}_\mathrm{tot}}\frac{\mathrm{d}\rho_\mathrm{GW}(\eta,k)}{\mathrm{d}\ln k} = \frac{1}{24}\left(\frac{k}{\calH(\eta)}\right)^{2}\overline{\mathcal{P}^{(s)}_h(\eta,k)}.
\eeq
 Finally, the GW spectral density $\Omega_\mathrm{GW}$ at PBH formation time, namely at horizon crossing time during the HBB expanding phase, will be given by~\cite{Kohri:2018awv}
\beq\label{eq:Omega_GW_f}
\begin{split}
\Omega_\mathrm{GW}(\eta_\mathrm{f},k)  & = \frac{1}{12}\int_{0}^{\infty} \mathrm{d}v\int_{|1-v|}^{1+v}\mathrm{d}u \left[ \frac{4v^2 - (1+v^2-u^2)^2}{4uv}\right]^{2}\\ & \times \mathcal{P}_\mathcal{R}(kv)\mathcal{P}_\mathcal{R}(ku)  \left[\frac{3(u^2+v^2-3)}{4u^3v^3}\right]^{2} \\ & \times \biggl\{\biggl[-4uv + (u^2+v^2-3)\ln \left| \frac{3 - (u+v)^{2}}{3-(u-v)^{2}}\right|\biggr]^2  \\ & + \pi^2(u^2+v^2-3)^2\Theta(v+u-\sqrt{3})\biggr\}.
\end{split}
\eeq
Lastly, considering the entropy conservation between PBH formation time and the present epoch, we can show that
\beq\label{Omega_GW_RD_0}
\Omega_\mathrm{GW}(\eta_0,k) = \Omega^{(0)}_r\frac{g_{*\mathrm{\rho},\mathrm{f}}}{g_{*\mathrm{\rho},0}}\left(\frac{g_{*\mathrm{S},\mathrm{0}}}{g_{*\mathrm{S},\mathrm{f}}}\right)^{4/3}\OmegaGW(\eta_\mathrm{f},k),
\eeq
where the subscript $0$ refers to the present epoch and $g_{*\mathrm{\rho}}$ and $g_{*\mathrm{S}}$ denote the energy and entropy relativistic degrees of freedom. For our  numerical applications  we use $\Omega_\mathrm{rad,0}\simeq 10^{-4}$~\cite{Planck:2018vyg}, $g_{*\mathrm{\rho},0}\simeq g_{*\mathrm{S},0}= 3.36$, $g_{*\mathrm{\rho},\mathrm{f}}\simeq g_{*\mathrm{S},\mathrm{f}} = 106.75$~\cite{Kolb:1990vq}.

\begin{figure*}[h!]
\begin{center}
\includegraphics[width=0.796\textwidth]{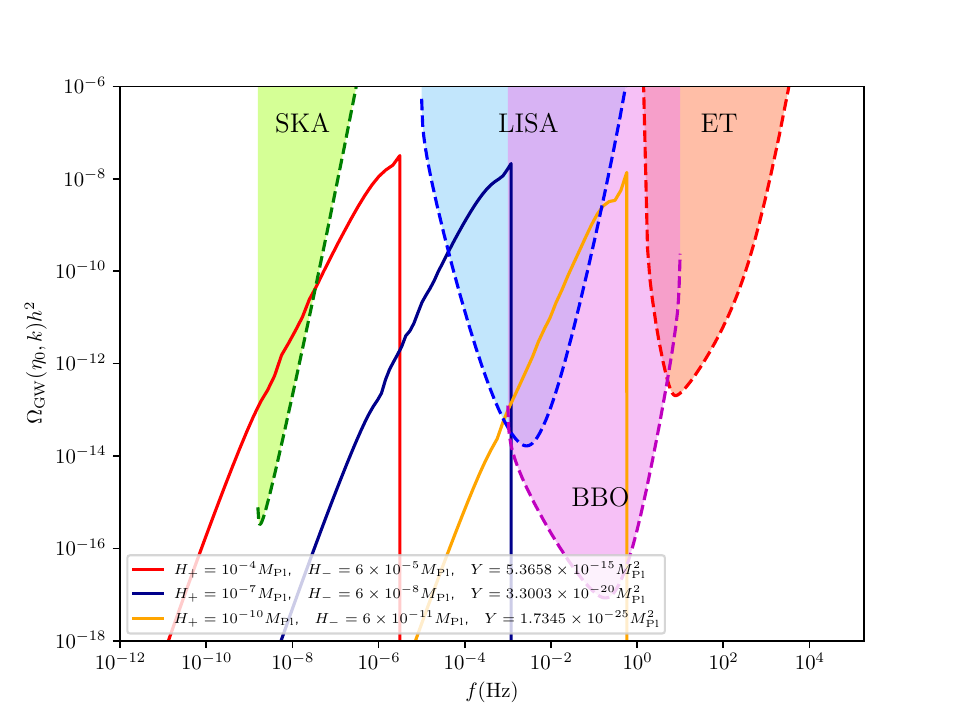}
\caption{{\it{The scalar-induced gravitational-wave spectra for different values of the parameters $H_{+}$, $H_{-}$ and $\Upsilon$. On   top of the   GW spectra we present   the sensitivity curves of SKA~\cite{Janssen:2014dka}, LISA~\cite{LISACosmologyWorkingGroup:2022jok}, BBO~\cite{Harry:2006fi} and ET~\cite{Maggiore:2019uih} GW experiments.}}}
\label{fig:Omega_GW}
\end{center}
\end{figure*}

In \Fig{fig:Omega_GW}  we present the current GW spectral abundance  as a function of the frequency $f$ defined as $f\equiv \frac{k}{2\pi a_0}$, for different sets of our parameters at hand, namely $H_+$, $H_-$, and $\Upsilon$. Furthermore, we superimpose  the GW sensitivity bands of the forthcoming  GW experiments, namely   Square Kilometer Arrays (SKA)~\cite{Janssen:2014dka}, Laser Inferometer Space Antenna (LISA)~\cite{LISACosmologyWorkingGroup:2022jok}, Big Bang Observer (BBO)~\cite{Harry:2006fi} and Einstein Telescope (ET)~\cite{Maggiore:2019uih}. As one can see, at first $\Omega_\mathrm{GW}\propto f^2$ and then it decays abruptly at an ultra-violet (UV) cut-off frequency $f_\mathrm{UV} = \frac{2k_\mathrm{NL}}{2\pi a_0}$ related to the non-linear cutoff introduced in \Sec{sec:bounce} where $\mathcal{P}_\mathcal{R}(k_\mathrm{NL}) = 0.1$. Beyond this non-linear cut-off frequency  perturbation theory breaks down, and one needs to perform numerical simulations in order to derive the GW spectral behaviour in these high frequencies~\cite{Fernandez:2023ddy}, an investigation that goes beyond the scope of the present work.

The scaling $f^2$ of the GW spectral abundance in the low frequency range  can be seen from \Eq{eq:Omega_GW_f}, where $\Omega_\mathrm{GW}\propto \mathcal{P}^2_\mathcal{R}$. Thus, since $\mathcal{P}_\mathcal{R}\propto k$ (see \Eq{eq:P_R_anal}), we obtain $\Omega_\mathrm{GW}\propto k^2 \propto f^2$. Moreover, it is worth noticing    that the UV cut-off frequency at $2k_\mathrm{NL}/(2\pi a_0)$ can be justified by the momentum conservation, since as it can be seen by \Eq{Tensor Power Spectrum}  the tensor power spectrum $\mathcal{P}_h(k)$ is actually a convolution product of the curvature power spectrum $\mathcal{P}_\mathcal{R}(k)$, i.e. two scalar modes $\mathcal{R}$ give a tensor mode $h$. This explains the factor of $2$ in the UV cut-off frequency $f_\mathrm{UV}$. 

\begin{figure*}[h!]
\begin{center}
\includegraphics[height=6.cm,width=7cm]{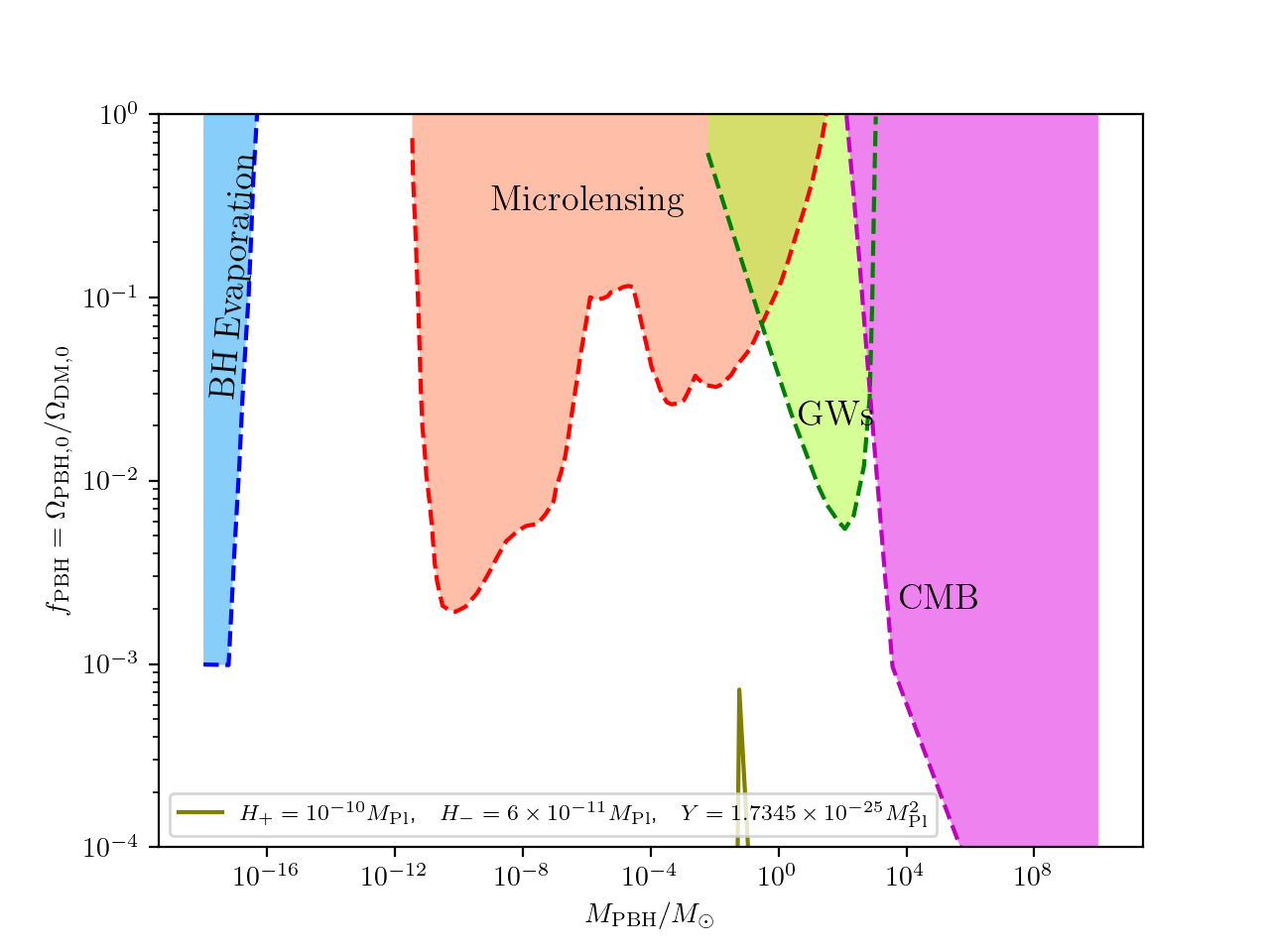}
\includegraphics[height=6.cm,width=7cm]{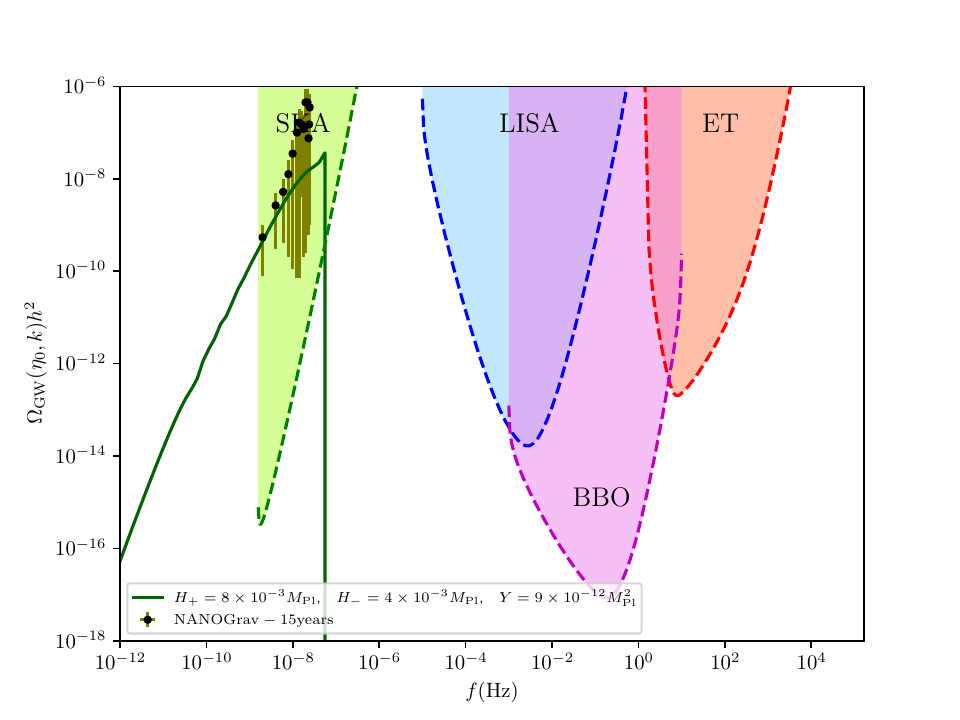}
\caption{{\it{Left Panel:  The fraction of dark matter in terms of PBHs denoted as $f_\mathrm{PBH}=\Omega_\mathrm{PBH,0}/\Omega_\mathrm{DM,0}$ as a function of the PBH mass for $H_+=8\times 10^{-3}\Mp$, $H_-=4\times 10^{-3}\Mp$ and $\Upsilon = 9\times 10^{-12}\Mp^2$. The colored regions are excluded from evaporation (blue region), microlensing (red region), gravitational-wave (green region) and CMB (violet region) observational probes concerning the PBH abundances. The data for the constraints on $f_\mathrm{PBH}$ from the different observational probes were obtained from~\cite{Green:2020jor}. Right Panel: The scalar-induced gravitational-wave spectrum for $H_+=8\times 10^{-3}\Mp$, $H_-=4\times 10^{-3}\Mp$ and $\Upsilon = 9\times 10^{-12}\Mp^2$, in comparison with the NANOGrav GW data  \cite{NANOGrav:2023gor}. On   top of our GW spectra we additionally present  the sensitivity curves of SKA~\cite{Janssen:2014dka}, LISA~\cite{LISACosmologyWorkingGroup:2022jok}, BBO~\cite{Harry:2006fi} and ET~\cite{Maggiore:2019uih} GW experiments.}}}
\label{fig:Omega_GW_NANOGrav}
\end{center}
\end{figure*}

In the right panel of \Fig{fig:Omega_GW_NANOGrav}  we show   the GW spectral abundance as a function of the frequency, for $H_+ = 8\times 10^{-3}\Mp$, $H_- = 4\times 10^{-3}\Mp$ and $\Upsilon = 9 \times 10^{-12}\Mp^2$, superimposed with the recently Pulsar Time Array (PTA) GW data released by NANOGrav  \cite{NANOGrav:2023gor}. As one may see, our GW prediction for the fiducial values of $H_+$, $H_-$ and $\Upsilon$ reported above, peaks at $\mathrm{nHz}$ and it can explain quite well the PTA GW data. Hence, it indicates that  the non-singular bouncing cosmological induced GW portal can serve as one of the possible interpretations for the NANOGrav/PTA GW signal~\footnote{Note here that the scalar-induced GW scenario related to PBH formation has been also extensively studied as a possible interpretation of the NANOGrav/PTA GW data within other than bouncing cosmological setups~\cite{Franciolini:2023pbf,Cai:2023dls,Balaji:2023ehk,Wang:2023ost,Yi:2023mbm,Basilakos:2023xof,Bhaumik:2023wmw,Choudhury:2023fwk,Choudhury:2023fjs}.}. A more careful likelihood analysis is needed in order to find the $H_+$, $H_-$ and $\Upsilon$ values best fitting the NANOGrav/PTA GW data at $\mathrm{nHz}$. Such an analysis is going beyond of the scope of the present work and it will be performed elsewhere. For consistency, we show in the left panel of \Fig{fig:Omega_GW_NANOGrav}, the contribution of PBHs to dark matter for $H_+ = 8\times 10^{-3}\Mp$, $H_- = 4\times 10^{-3}\Mp$ and $\Upsilon = 9 \times 10^{-12}\Mp^2$, showing that we do not face a PBH overproduction issue, being compatible with the PBH constraints.

\section{Conclusions}\label{sec:Conclusions}

The non-singular bouncing cosmological paradigm, being an attractive alternative to inflation, is free of the initial singularity problem,  being additionally able  to address the current HBB cosmological issues, namely the horizon and the flatness problems. Moreover, it is compatible with the CMB and LSS observational data, indicating a scale-invariant curvature power spectrum on large scales. 

Interestingly, PBHs  can serve as a novel portal in order to probe alternative cosmological and gravitational scenarios. Notably, in this work  we found a novel natural model-independent mechanism for PBH formation during the HBB radiation-dominated era, within the context of non-singular matter bouncing cosmologies.
In particular, the enhancement of super-horizon curvature perturbations, during a matter contracting phase in combination with a short transitory period from the matter contracting to the HBB expanding Universe, can lead to enhanced curvature perturbations on small scales during the HBB phase, collapsing to form PBHs.

Remarkably, the PBHs produced within our model-independent bouncing setup can lie within a wide range of masses, depending on the energy scales  at the end of the contracting era $H_-$, and at the beginning of the HBB expanding era $H_+$, as well as on the rate of growth of the Hubble parameter during the bouncing phase $\Upsilon$. Intriguingly, for $H_+ = 10^{-10}\Mp$, $H_- = 6\times 10^{-11}\Mp$ and $\Upsilon = 1.7345\times 10^{-25}\Mp^2$, we find  PBHs lying within the observationally unconstrained asteroid-mass window, where PBHs can  potentially account for the totality of dark matter. 

Furthermore, we  studied  the stochastic GW background, induced by second order gravitational interactions and by the enhanced curvature perturbations collapsing to PBHs. Interestingly, we   found an abundant production of induced GWs, peaking at a frequency ranging from $\mathrm{nHz}$ up to $\mathrm{Hz}$, depending on the value of $H_+$, $H_-$ and $\Upsilon$, hence being potentially detectable by future GW experiments, in particular  SKA, PTAs, LISA and ET, and serving as a novel probe of the potential bouncing nature of  initial conditions prevailing in the early Universe. Lastly, we showed that our non-singular bouncing setup can give rise to a stochastic induced GW background peaked at $\mathrm{nHz}$, being able to explain quite efficiently the recently released PTA/NANOGrav GW data.

\begin{acknowledgments}
  TP and SC  acknowledge the support of the INFN Sezione di Napoli \textit{initiativa specifica} QGSKY. TP and ENS acknowledge the  contribution of the LISA Cosmology Working Group. SC, TP and ENS acknowledge the  contribution of the COST Action CA21136 ``Addressing observational tensions in cosmology with systematics and  fundamental physics (CosmoVerse)''. TP  acknowledges as well financial support from the Foundation for Education and European Culture in Greece.
YFC is supported in part by the National Key R\&D Program of China (2021YFC2203100), CAS Young Interdisciplinary Innovation Team (JCTD-2022-20), NSFC (12261131497), 111 Project (B23042), CSC Innovation Talent Funds, USTC Fellowship for International Cooperation, USTC Research Funds of the Double First-Class Initiative. 
\end{acknowledgments}

\bibliographystyle{JHEP} 
\bibliography{PBH}

\end{document}